\documentclass[useAMS,fleqn,usenatbib]{mn2e}
\usepackage{times}
\usepackage{amsmath}
\usepackage{bm}
\usepackage{url}
\usepackage{graphicx}
\usepackage{hyperref}
\usepackage{float}
\input amssym.tex

\mathchardef\mhyphen="2D

\newcommand{\PDF}{\mathcal{P}}

\newcommand{\hmpc}{\,$h^{-1}$\,Mpc}

\newcommand{\hmpcnosp}{$h^{-1}$\,Mpc}

\newcommand{\lcdm}{$\Lambda$CDM}

\newcommand{\deltal}{\delta_\ell}

\newcommand{\deltac}{\delta_c}
\newcommand{\deltallin}{\delta_\ell^{\rm lin}}
\newcommand{\deltaslin}{\delta_s^{\rm lin}}
\newcommand{\sigmal}{\sigma_\ell}
\newcommand{\sigmas}{\sigma_s}

\newcommand{\kcut}{k_{\rm cut}}
\newcommand{\rcut}{R_{\rm cut}}
\newcommand{\vs}{\nonumber\\} 

\chardef\til=`\~

\begin{document}

\title[Halo bias in voids]{A halo bias function measured deeply into voids without stochasticity}

\author[Neyrinck et al.]
{Mark C.\ Neyrinck$^1$\thanks{E-mail:neyrinck@pha.jhu.edu}, Miguel A. Arag\'{o}n-Calvo$^{1,2}$, Donghui Jeong$^1$, Xin Wang$^1$\\
$^1$Department of Physics and Astronomy, The Johns Hopkins University, Baltimore, MD 21218, USA\\
$^2$Department of Physics and Astronomy, University of California, Riverside, CA 92521, USA}

\maketitle

\begin{abstract}
We study the relationship between dark-matter haloes and matter in the MIP $N$-body simulation ensemble, which allows precision measurements of this relationship, even deeply into voids. What enables this is a lack of discreteness, stochasticity, and exclusion, achieved by averaging over hundreds of possible sets of initial small-scale modes, while holding fixed large-scale modes that give the cosmic web.  We find (i) that dark-matter-halo formation is greatly suppressed in voids; there is an exponential downturn at low densities in the otherwise power-law matter-to-halo density bias function. Thus, the rarity of haloes in voids is akin to the rarity of the largest clusters, and their abundance is quite sensitive to cosmological parameters.  The exponential downturn appears both in an excursion-set model, and in a model in which fluctuations evolve in voids as in an open universe with an effective $\Omega_m$ proportional to a large-scale density.  We also find that (ii) haloes typically populate the average halo-density field in a super-Poisson way, i.e.\ with a variance exceeding the mean; and (iii) the rank-order-Gaussianized halo and dark-matter fields are impressively similar in Fourier space.  We compare both their power spectra and cross-correlation, supporting the conclusion that one is roughly a strictly-increasing mapping of the other. The MIP ensemble especially reveals how halo abundance varies with `environmental' quantities beyond the local matter density; (iv) we find a visual suggestion that at fixed matter density, filaments are more populated by haloes than clusters.
\end{abstract}

\begin {keywords}
  large-scale structure of Universe -- cosmology: theory
\end {keywords}
\section{Introduction}
The spatial arrangement of matter and galaxies contains a large fraction of the information available about cosmology and galaxy formation, especially at late times. The dynamics of collisionless dark matter is straightforward to model in simulations, but the distribution of the easiest-to-observe tracers of the dark matter, the galaxies, may have a complicated relationship to the distribution of dark matter. 

The mapping from dark matter to galaxies is often treated in two parts: the mapping from the dark-matter field to haloes, and from haloes to galaxies. The dark-matter-to-halo mapping is rather straightforward, since it only involves dark-matter physics. The halo-to-galaxy mapping involves baryonic physics, so may be quite complicated.  Thankfully, the relationship seems to be simpler than it could be in principle. Relatively large galaxies seem to relate quite well to dark-matter subhaloes, through subhalo abundance matching \citep[e.g.][]{BrainerdVillumsen1994,KravtsovKlypin1999,NeyrinckEtal2004,KravtsovEtal2004,ConroyEtal2006}. Another common strategy is to populate galaxies in haloes according to a halo occupation distribution \citep[HOD, e.g.\ ][]{BerlindWeinberg2002}.

Inferring the presence of haloes and subhaloes from dark matter can be done by (sub)halo finding in $N$-body simulations \citep[e.g. ][]{KnebeEtal2013}.  While this relationship is conceptually straightforward, it is useful to model the process using analytical approximations as well, for example to simply interpret observed galaxy clustering. A nontrivial relationship between matter and galaxies is called bias \citep[e.g.][]{Kaiser1984,MoWhite1996,ManeraGaztanaga2011,ParanjapeEtal2013}.

In this paper, we investigate the approximation that matter and halo density fields are related by a local monotonic bias function on few-Mpc scales, that may be non-linear.  This formulation in terms of a general biasing function is an idea that has been studied \citep[e.g.][]{Szalay1988,Matsubara1995,SigadEtal2000,Matsubara2011,FruscianteSheth2012} but current work in this area tends to formulate the bias in terms of the first few Taylor-series coefficients. Indeed, on large, linear scales, the bias takes an approximately linear form when measured through the power spectrum, i.e. $P_g(k)=b^2P_m(k)$, for a scale-independent bias $b$. Here, $P_g$ denotes the power spectrum of the galaxy overdensity field $\delta_h=(\rho_h/\bar{\rho}_h)-1$, and $P_m$ is the power spectrum of the matter overdensity field $\delta_m$. The assumption underlying this is that the halo overdensity is a constant $b$ times the matter overdensity, i.e.\ $\delta_h=b\delta_m$, which is a good approximation on large scales. But this relationship is unphysical if it is assumed that it holds literally for $b\ne1$, down to the lowest densities, since where there is no matter ($\delta_m=-1$), a linear bias with $b\ne 1$ predicts that $\delta_h$ will have positive or even negative density. 

Parameterizing the bias in terms of log-densities, for example letting $\ln(1+\delta_h)=b\ln(1+\delta_m)$, avoids these issues, giving something that in principle could hold literally. Various authors have used a power law in $(1+\delta)$ \citep{CenOstriker1993,delaTorrePeacock2013,KitauraEtal2014}, which is equivalent to a first-order bias in  the log-density variable. Indeed, \citet{JeeEtal2012} found that formulating bias explicitly in terms of log-densities, and weighting haloes by their masses, results in a two-second-order bias function with low scatter. Log-densities are natural density variables \citep{ColesJones1991}, and their statistics are generally better-behaved than those of the overdensity $\delta$ \citep{NeyrinckEtal2009,Carron2011}.

Using the local bias-function framework, a mock halo catalog can be produced from a low-resolution (with e.g.\ few-Mpc cells) matter density field $\delta_m$ by mapping $\delta_m$ to a `continuous' halo-density field $\delta_h$. This continuous field can then be sampled into discrete haloes, for example in a Poisson fashion. Recently, \citet{KitauraEtal2014} showed that this approach is useful for producing fast mock halo catalogs that have halo power spectra consistent with those from full $N$-body simulations \citep[see also][]{ChanScoccimarro2012}. They use Augmented Lagrangian Perturbation Theory \citep[ALPT;][]{KitauraHess2013} to produce $N$-body realizations that are accurate on few-Megaparsec scales. ALPT interpolates in Fourier space between second-order Lagrangian perturbation theory on large scales, and a spherical-collapse approach \citep{Neyrinck2012} on small scales, to fix second-order Lagrangian perturbation theory's problems at high and low densities. This approach will be useful, for instance, to produce the vast number of simulations necessary for current and upcoming surveys to get sufficiently accurate covariance matrices for precision cosmological constraints (e.g.\ from baryon acoustic oscillations).

To investigate the bias function, we use the MIP ({\it Multum In Parvo}, `many things in the same place') ensemble of simulations \citep{AragonCalvo2012}. We use 225 simulations from this suite, all of which have the same large-scale cosmic web (built from modes with initial modes of wavelength $2\pi/k\ge4$\hmpc). Each simulation has a different set of initial small-scale modes (with wavelength $<4$\hmpc).  The 4\hmpc\ scale roughly divides two regimes in scale: those giving the dominant components of the cosmic web, and those giving finer details \citep{BondEtal1996}. See for example \citet{SuhhonenkoEtal2011} for an examination of how different scale regimes contribute to the cosmic web, using $N$-body simulations. The 4\hmpc\ dividing scale is a bit arbitrary within a factor  $<2$: the linear-theory density variance in spheres of radius $R$, $\sigma^2_R$, crosses unity between $R=4$\hmpc\ and $R=8$\hmpc. This means that some stream crossing (e.g.\ as produced in the Zel'dovich approximation) occurs on larger scales than 4\hmpc, giving a cosmic web, but only the `top level' of it, not the full cosmic-web hierarchy \citep{AragonCalvoSzalay2013}. This scale boundary was also guided by eye: simulations were run with power zeroed out below a few different scale thresholds.  Zeroing out power of wavelength $<4$\hmpc\ still gives a visually evident cosmic web, but not if the threshold is increased to 8.  This ensemble enables a sort of `cosmic-web occupation distribution,' giving a set of possible ways each location in the `top-level' cosmic web might be occupied by different types of haloes.

The MIP-ensemble mean gives an estimate of the continuous halo-density field $\delta_h$ with negligible discreteness, and also largely free of halo-halo exclusion effects.  Halo exclusion may suppress the halo density in clusters, but in the stacked ensemble, haloes may be arbitrarily close to each other, unlike in a single simulation.

Halo-bias stochasticity \citep[e.g.\ ][]{Pen1998, TegmarkBromley1999, DekelLahav1999, Matsubara1999, ShethLemson1999, TaruyaSoda1999, SeljakWarren2004, NeyrinckEtal2005,HamausEtal2010,BaldaufEtal2013} is something that the MIP ensemble offers an ideal laboratory to explore, at least with stochasticity defined in a particular way.  There are two types of stochasticity present. First, there are fluctuations in the MIP ensemble-mean `continuous' $\delta_h$ away from that predicted from the matter field $\delta_m$ using a deterministic biasing function, $\delta_h(\delta_m)$. These fluctuations are from environmental effects beyond $\delta_m$. Second, on top of that, there is a sampling stochasticity, in the way the continuous $\delta_h$ gets point-sampled in each realization to get the halo sample. This second, sampling stochasticity is imparted by MIP realization-to-realization changes in small-scale modes.

Voids are emerging as potentially powerful cosmological tools \citep[e.g.][]{Ryden1995,GranettEtal2008,BiswasEtal2010,LavauxWandelt2012,BosEtal2012,LiEtal2012,SpolyarEtal2013,HamausEtal2014}. For many of these techniques, precision constraints will require precision knowledge of the relationship between haloes and matter within voids.  This relationship has already seen much study \citep{Peebles2001,GottloberEtal2003,FurlanettoPiran2006,TinkerConroy2009,JenningsEtal2013,SutterEtal2013}, but the MIP ensemble allows us to measure it with precision to unprecedentedly low densities.

This paper is laid out as follows. In $\S$\ref{sec:models} we describe two analytic models for the relationship between halo and matter densities. These include an `additive'-excursion-set model in $\S$\ref{sec:aes}, and a local-growth-factor model, in which the effective growth factor of small fluctuations is given by the local large-scale density.  In $\S$\ref{sec:mattertohaloes}, we show measurements of this from the MIP ensemble. In $\S$\ref{sec:poissonity}, we test the assumption that the halo population in a given realization is a simple Poisson sampling of the continuous halo-density field. In $\S$\ref{sec:cosmicweb}, we visually assess factors besides the density that influence the halo-matter bias relationship. Finally, in $\S$\ref{sec:gaussianization}, we compare the rank-order-Gaussianized $\delta_m$ and $\delta_h$ fields in Fourier space. If one is a strictly increasing function of the other, they should be equal after Gaussianizing each of them to the same probability density function (PDF). 

\section{Models of the bias function}
\label{sec:models}
In this section we discuss two models for the dark-matter-to-halo relationship; casual readers may wish to skip to the next section, which shows our results.  We explore two approaches: (1) an `additive' excursion-set (AES) approach in which small-scale modes simply contribute additively to a large-scale density as they attempt to draw mass over the spherical-collapse barrier for halo formation; and (2) a new local-growth-factor (LGF) model, in which halo formation on small scales proceeds as though it were in a homogeneous FRW universe with an effective $\Omega_M^{\rm eff}$ depending on the large-scale density.

In the Press-Schechter model \citep{PressSchechter1974}, the average number density of haloes of mass $M$ is
\begin{equation}
n(M) = \sqrt{\frac{2}{\pi}}\frac{\bar{\rho}}{M^2}
\left|\frac{d\ln\sigma(M)}{d\ln M}\right|
\frac{\deltac}{\sigma(M)}\exp\left(-\frac{\deltac^2}{2\sigma^2(M)}\right).
\label{eqn:nmps}
\end{equation}
Here, $\sigma^2(M)$, an integral over the power spectrum, is the variance in the linear-theory density field in a top-hat sphere encompassing mass $M$ in Lagrangian space (of radius $R(M)$), and $\deltac=1.686$ is the threshold in the linearly-extrapolated density for spherical collapse.

The Press-Schechter model is known to have shortcomings in halo mass-function predictions; the sharp-$k$ filter used for the random walk \citep{BondEtal1991} is arguably unnatural, although it simplifies calculations. Also, the collapse or expansion is not generally spherical.  Departures from these assumptions can be modelled \citep[e.g.][]{ShethEtal2001,CorasanitiAchitouv2011,AchitouvEtal2013}, although unmodelled complexities in halo formation may persist in any simple model \citep{LudlowPorciani2011}. 

\subsection{Additive excursion-set model}
\label{sec:aes}
While the Press-Schechter model is usually thought to be oversimplified, the MIP ensemble is unusually suited to a sharp-$k$ random-walk excursion-set model \citep{BondEtal1991,MoWhite1996}, since there is a decoupling between long and short modes, with a sharp-$k$ boundary at $\kcut=2\pi/\rcut$, where $\rcut=4$\hmpc. Denote the linearly-extrapolated Lagrangian initial density at a given location as $\delta^{\rm lin}$. We split this into long and short (or large-scale and small-scale) components, $\deltallin$ (the same in all MIP realizations) and $\deltaslin$ (different in each realization). 

Our measurements of $\delta_h$ are as a function of the non-linear density $\deltal$ as measured in the simulation.  But the excursion-set model requires linear-theory densities, so we use a one-to-one mapping giving $\deltallin(\deltal)$ in the spherical-shell evolution model, given parametrically e.g.\ by \citet{ShethvandeWeygaert2003}.

Conceptually, holding $\deltal$ fixed changes Eq.\ (\ref{eqn:nmps}) in a few ways. It changes the barrier for collapse, $\deltac\to(\deltac-\deltallin)$. It also reduces the variance of fluctuations from which haloes can form, $\sigma^2(M)\to\sigmas^2(M)$, where $\sigmas^2(M)$ is the variance of a top-hat sphere applied to $\deltaslin$, i.e.\ zeroing out modes with $k<k_{\rm cut}$.  Also defining $\sigmal^2$ to be the variance of $\deltallin$, if $R(M)\ll \rcut$, then $\sigma^2(M)=\sigmal^2+\sigmas^2(M)$. [$R(M)\ll \rcut$ ensures that the difference between the sharp-$k$ filter used for $\sigmal$ and the sharp-$x$ filter used for the other two terms is negligible for $k<\kcut$.] There is another change: since the Press-Schechter model is really a Lagrangian model, there is an added factor in the conversion to Eulerian space, $(1+\deltal)$ (implicitly there before, with $\deltal=0$). The result can be explicitly calculated from the cumulative distribution,
\begin{align}
&n(M|\deltal)=(1+\deltal)n_{\rm Lagrangian}(M|\deltal)
\vs
=&
-(1+\deltal)\frac{\bar{\rho}}{M}\frac{d}{dM} 
\left[
\mathrm{erfc}\left(
\frac{\deltac - \deltallin}{\sqrt{2(\sigma^2(M)-\sigmal^2)}}
\right)
\right]
\vs
=&
-(1+\deltal)\frac{\bar{\rho}}{M}\frac{d\sigma(M)}{dM}
\frac{d}{d\sigma(M)}\left[
\mathrm{erfc}\left(
\frac{\deltac - \deltallin}{\sqrt{2\sigmas^2(M)}}
\right)
\right]
\vs
=\ &
(1+\deltal)\sqrt{\frac{2}{\pi}}\frac{\bar{\rho}}{M^2}\frac{\sigma^2(M)}{\sigmas^3(M)}
\left|\frac{d\ln\sigma_M}{d\ln M}\right|
\frac{\deltac - \deltallin}{\deltac}\times\vs
& \exp\left[
-\frac{(\deltac - \deltallin)^2}{2\sigmas^2}
\right].
\label{eqn:nmpsdelta}
\end{align}
\citet{FurlanettoPiran2006} give the same expression for $n_{\rm Lagrangian}$.
 
Note that there is also extra factor of $\sigma^2(M)/\sigmas^2(M)$ compared to Eq.\ (\ref{eqn:nmps}), beyond the conceptual substitutions discussed above.  In Lagrangian space, this ensures that $\int n_{\rm Lag}(M | \deltallin) \PDF(\deltallin) d\deltallin = n(M)$, using the Gaussian distribution of $\deltallin$, $\PDF(\deltallin)=\exp[-(\deltallin/\sigmal)^2/2]/\sqrt{2\pi\sigmal^2}$. Note that here we include $\deltallin>\delta_c$ in the range of integration, which arguably should not be done since patches of this initial density have collapsed \citep{ShethLemson1999,MussoEtal2012}. Accounting for this would change the normalization, but negligibly in our case, since $\sigmal\approx 1$; integrating a Gaussian of unit variance up to 1.686 gives $>99$\% of the total area.

Dividing Eq.\ (\ref{eqn:nmpsdelta}) by Eq.\ (\ref{eqn:nmps}) gives
\begin{align}
1+\delta_h(M | \deltal) =\ &(1+\deltal)\frac{\sigma^3(M)}{\sigmas^3(M)}\frac{\deltac-\deltallin}{\deltac}\times\nonumber \\
& \exp\left\{-\frac{1}{2}\left[\left(\frac{\deltac-\deltallin}{\sigma_s(M)}\right)^2 - \frac{\deltac^2}{\sigma^2(M)} \right]\right\},
\label{eqn:aes}
\end{align}
or, slightly more simply, in log-density variables $A\equiv \ln(1+\delta)$,
\begin{align}
A_h(M | A_\ell) =\ &A_\ell + \ln\left[\frac{\sigma^3(M)}{\sigmas^3(M)}\frac{\deltac-\deltallin}{\deltac}\right] - \\
& \frac{1}{2}\left[\left(\frac{\deltac-\deltallin}{\sigma_s(M)}\right)^2 - \frac{\deltac^2}{\sigma^2(M)} \right].
\label{eqn:aesa}
\end{align}

Note that in the MIP ensemble, there is a difference between $\deltal$ and the matter density measured in an Eulerian cell, the `void density' $\delta_v$, in the notation of \citep{FurlanettoPiran2006}. This is because in the MIP, $\deltal$ truly comes from a sharp-$k$ cut in Lagrangian space.  In a usual simulation, or in reality, there would be no such Lagrangian cut, and $\deltal$ as estimated on a grid would be filtered through a Eulerian pixel window function.  For the comparison below in Fig.\ \ref{fig:biasscatter}, there are two windows being applied to obtain $\delta_m$: a Lagrangian sharp-$k$ filter, and the Eulerian pixel window function. In comparing the model to the measurements, we are neglecting the Eulerian pixel window function, which is likely negligible compared to the Lagrangian filter. In applying Eq.\ (\ref{eqn:aes}) outside the MIP, we suspect that in calculating $\sigmas^2(M)$, it would work adequately to filter the linear power spectrum with the Eulerian pixel window function.

\subsection{Local-growth-factor model}
\label{sec:lgf}
In this section, we describe a model in which fluctuations in halo number density involve a local growth factor (LGF), $D(\deltal)$, that depends on the large-scale density $\deltal$.  For instance, by Birkhoff's theorem, inside a spherically symmetric void, the dynamics inside can be treated like an FRW universe with a modified $\Omega_m$ \citep[e.g.][]{ShethvandeWeygaert2003}; this concept has rather wide applicability \citep[e.g.][]{BaldaufEtal2011,SherwinZaldarriaga2012}. We modify the growth factor according to a scaled $\Omega_m$. As \citet{MartinoSheth2009} show, this model should give the same results as the AES model if there is also a local, effective $\Omega_\Lambda$, modified to account for local fluctuations to the Hubble constant \citep{GoldbergVogeley2004}. We do not implement this effective $\Omega_\Lambda$ change, for simplicity, since this effective $\Omega_\Lambda$ is solved for numerically.  As we will see below, there seems to be a rather good agreement between the models at low density, even without the additional change.

As before, we work with the small-scale linear density field $\deltaslin$, but we assume that it gets amplified by a factor $D(\deltal)/D_0$. Here, $D_0$ is the global growth factor, and $D(\deltal)$ is a local growth factor, estimated using $\Omega_m^{\rm eff} = \Omega_m(1+\deltal)$, and with unchanged $\Omega_\Lambda$. Note that one must use the non-linear $\deltal$ here; using $\deltallin$ can give nonsense in voids, since $1+\deltallin$ is not constrained to be positive.  For calculations, we used the {\scshape grow$\lambda$} package \citep{Hamilton2001}. We also tried using the expression \citep{LahavEtal1991}
\begin{equation}
D(\deltal) = \frac{5}{2}\frac{a\Omega_m^{\rm eff}}{(\Omega_m^{\rm eff})^{4/7}-\Omega_\Lambda+(1+\Omega_m^{\rm eff}/2)(1+\Omega_\Lambda/70)};
\label{eqn:lahav}
\end{equation}
it gives very similar results, but there are slight visible differences. As \citet{Hamilton2001} states, all of these growth-factor formulae are invalid for $\Omega_m$ sufficiently large to give a collapsing universe, so the results in the LGF model for large $\deltal$ should be used with caution.

Compared to Eq.\ (\ref{eqn:nmps}), we make the following changes: $\sigma(M)\to\sigmas(M)D(\deltal)/D_0$; and we additionally multiply the whole expression by the Lagrangian-to-Eulerian factor $(1+\deltal)$. This gives 
\begin{align}
 n(M | \deltal) =\  &  
  (\deltal + 1) \sqrt{\frac{2}{\pi}}\frac{\bar{\rho}}{M^2}
\left|\frac{d\ln\sigma(M)}{d\ln M}\right|
 \frac{\deltac}{\sigmas(M)}\frac{D_0}{D(\deltal)}  \nonumber\\
 & \times \exp\left[-\frac{1}{2}\left(\frac{\deltac}{\sigmas(M)}\frac{D_0}{D(\deltal)}\right)^2\right].
 \label{eqn:nmdlgf}
 \end{align}
Arguably, the logarithmic derivative here should be changed to use $\sigmas(M)$, but the normalization is a bit uncertain anyway. [In the AES model, the logarithmic derivative is the same as here, but it contributed a factor of $\sigma(M)/\sigmas(M)$.] 
Eq.\ (\ref{eqn:nmdlgf}) is normalized to simply give $n(M | \deltal = 0) = n(M)$, i.e.\ giving curves that go through $(0,0)$. This does not happen in the AES model, and below in Fig.\ \ref{fig:biasscatter}, the measurements too seem to depart from $(0,0)$, most notably at high masses. Ideally, there would be a true normalization done, ensuring that $n(M) = \int n(M|\deltal) \PDF(\deltal)d\deltal$.  There are two reasons we do not work through this. First, this integral is over the non-linear $\PDF(\deltal)$, a simple, accurate analytic form for which we are not aware.  Second, as noted above, we do not expect the model to be accurate at high $\deltal$, so it is likely unwise to take any integral over the full range of $\deltal$ seriously. One strategy is to adopt the normalization from the AES model, which is what we do below. Also, the Eulerian excursion-set model introduced by \citet{Sheth1998} may be of use in determining the proper normalization. We leave a thorough investigation of this normalization to future work.

Dividing Eq.\ (\ref{eqn:nmdlgf}) by Eq.\ (\ref{eqn:nmps}) gives
\begin{align}
1+ \delta_h(M,\deltal) & =  (1+\deltal)\frac{D(\deltal)}{D_0} \times \nonumber \\
 & \exp\left[-\frac{1}{2}\left\{\left(\frac{\deltac}{\sigma_s(M)}\frac{D_0}{D(\deltal)}\right)^2 -
 \frac{\deltac^2}{\sigma^2(M)}\right\}\right].
\label{eqn:lgf}
\end{align}
This simplifies a bit in log-density variables:
\begin{equation}
A_h = A_m-\ln \frac{D(\deltal)}{D_0}  -\frac{1}{2}\left[\left(\frac{\deltac}{\sigma_s(M)}\frac{D_0}{D(\deltal)}\right)^2 -
 \frac{\deltac^2}{\sigma^2(M)}\right].
\label{eqn:lgfa}
\end{equation}

\section{The dark-matter-to-halo density relation}

\label{sec:mattertohaloes}
\begin{figure*}
  \begin{minipage}{\textwidth}
    \begin{center}
      \includegraphics[width=\columnwidth]{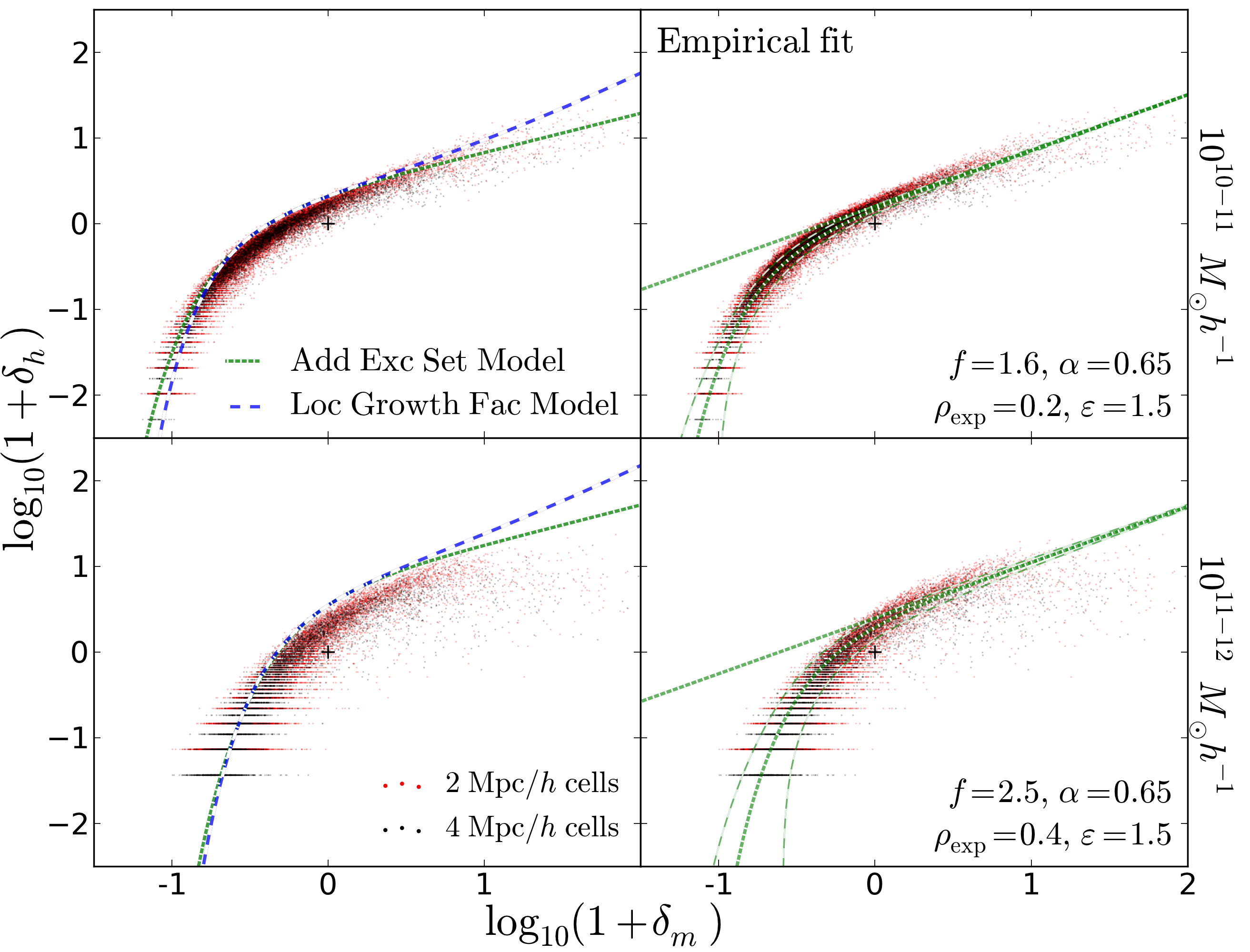}
    \end{center}
        \caption{Scatter plots of the MIP-ensemble-mean halo log-density versus the matter log-density, for two mass ranges of haloes. Each dot represents a 2\hmpc\ (red) or 4\hmpc\ (black) cubic grid cell, with an NGP-estimated density.  In the the right-hand column, empirical power-law-plus-exponential fits to the curve are shown in dotted curves, with and without the exponential.  The lighter, dashed curves show vertical one-$\sigma$ standard deviations in $\delta_h$ away from the mean, assuming Poisson statistics with mean given by the fit.  The left-hand column shows theory curves from the additive-excursion-set (AES) and local-growth-factor (LGF) models, Eqs.\ (\ref{eqn:aes}) and (\ref{eqn:lgf}). The black crosses indicate the origin, $(0,0)$.
        }
    \label{fig:biasscatter}
  \end{minipage}
\end{figure*}

Fig.\ \ref{fig:biasscatter} shows scatter plots of the halo and dark-matter densities in 2 and 4\hmpc\ grid cells in the ensemble-mean MIP fields at $z=0$.  The additive-excursion-set (AES) and local-growth-factor (LGF) models, described in the previous section, fit the curves rather well, especially at low density.

Each MIP simulation has 256$^3$ particles in a 32\hmpc\ box, and was run with vanilla \lcdm\ cosmological parameters: $\Omega_m=0.3$, $\Omega_\Lambda=0.7$, $h=0.73$, $\sigma_8=0.84$, and $n_s=0.93$. The particle mass is 1.6$\times10^8 M_\odot/h$; the haloes we analyze were found using a friends-of-friends (FoF) algorithm with linking length 0.2 times the Lagrangian particle spacing. The smallest haloes we use are of mass $10^{10} M_\odot/h$, i.e.\ consisting of at least 63 particles.  We did not use smaller-mass haloes than this because we noticed that the power spectrum of $10^{9.5\mhyphen10} M_\odot/h$ haloes (with 20-63 particles) had a higher amplitude than that of  $10^{10\mhyphen11} M_\odot/h$ haloes; this was likely from contamination in the $10^{9.5\mhyphen10} M_\odot/h$ bin with spurious haloes, bringing their power spectrum closer to the matter power spectrum (the $10^{10\mhyphen11} M_\odot/h$ haloes are under biased; see Fig.\ \ref{fig:gausspower} below).

It should be noted that the quite small, 32-\hmpcnosp\ box size implies that large-scale modes, and sheer volume, are missing that would substantially increase the range of environments encountered in the simulation \citep[e.g.][]{LukicEtal2007}.  The extreme clusters and voids in the ensemble are less extreme than one would find even in a random 32\hmpc\ volume of a large simulation. If a full sampling of cosmic-web environments were present, we suspect that the results on the low-density, void end would be nearly identical, just carried to smaller densities, because the environment deeply in a void is rather simple, always consisting of a single stream at low Lagrangian resolution. On the high-density end, however, our results are less robust, since in a large-volume simulation, a greater variety of cluster environments, merger histories, etc. would be present.

Density fields were measured using nearest-grid-point (NGP) assignment in each simulation, and then these density fields were averaged across the simulation ensemble.
A higher-order density-assignment scheme might give smoother plots, but we use NGP to make the level of halo discreteness in the plots obvious.  The particle discreteness is entirely negligible, with 225$\times16^3\approx10^6$ particles per cell on average in a 2\hmpc\ cell. 
The lowest-possible nonzero $(1+\delta_h)$ corresponds to only one halo in the grid cell, across all 225 MIP realizations.  We show results for haloes up to $\sim10^{12} M_\odot/h$, because we want the haloes analyzed to fit in a Lagrangian box of size 4\hmpc\ (and mass $\sim 7\times10^{12} M_\odot/h$) in the initial conditions.  This is important because we want to test for halo collapse from small-scale modes (below 4\hmpc), unmixed with larger-scale modes that are held fixed. 

The curves were calculated from the linear power spectrum using the {\scshape CosmoPy}\footnote{http://www.ifa.hawaii.edu/cosmopy/} package. To estimate $\sigma(M)$ in each case, we use halo masses of $2\times 10^{10}$ and $2\times10^{11} M_\odot/h$ approximately the median halo mass in each bin. 

Note that in translating the equations of $\S$\ref{sec:models} to this section, we use $\delta_m$ instead of $\delta_\ell$; here they both mean the matter density smoothed with 4-\hmpcnosp\ Lagrangian sharp-$k$ filter. The density pixelization adds an additional Eulerian 2- or 4-\hmpcnosp\ pixel window function, but this does little to the field that already has a 4-\hmpcnosp\ Lagrangian filter applied. Thus, it is not surprising that the 2- and 4-\hmpcnosp\ results line up with each other.

For simplicity and to highlight the similarity in shape between the two models, we have set the somewhat uncertain normalization of the LGF model similarly as in the AES model, adding a factor $[\sigma(M)/\sigmas(M)]^3$ [see Eq.\ (\ref{eqn:aes})] -- compared to what appears in Eq.\ (\ref{eqn:lgf}).  Note that this normalization does not go through $(0,0)$, the position of which on the plots is indicated by black crosses. The measured locus of points typically goes about halfway between $(0,0)$ and the predictions, so empirically, the best-fitting amplitude seems to be about the square root of the factor prescribed by the AES. The offset is the result of the zero-lag term in the local bias expansion \citep{SchmidtEtal2013}, which comes about because the mean number density of haloes in a finite region is different from the cosmic mean.

The AES and LGF curves have interestingly similar shapes at low density; according to \citet{MartinoSheth2009}, they should agree even better with an additional change to $\Omega_\Lambda$.  The LGF model seems to be a bit more accurate where the two curves diverge most in the low-density regime, at low halo mass (the top row). Changing the LGF normalization can change the quality of the fit, but perhaps not substantially on the low-density exponential tail. We examine the similarity between the AES and LGF models at low density a bit further in the Appendix. 

The success of the models at low densities has an interesting implication: the mass function of haloes within voids is highly sensitive to cosmology \citep{SongLee2009,Lee2012}.  Of course, here a `void' means a (matter) density depression \citep[e.g.][]{PlatenEtal2007,zobov}, not a structure entirely devoid of galaxies, in which, by definition, there would be no galaxies. In an effective low-$\Omega_m$ universe deeply in a void, the high-mass cutoff in the global mass function moves to lower masses \citep{GottloberEtal2003}. This makes the presence of even a globally modest-mass halo deep in a void possibly as rare as the highest-mass clusters in the Universe. Void-galaxy abundances could be used together with cluster abundances to test for primordial non-Gaussianity, as well, since both of them probe fluctuations in different large-scale density regimes.

Cosmological inference from void galaxies presents a few difficulties: baryonic processes are unlikely to be negligible in determining whether void haloes become populated with galaxies; and void-halo masses are likely even harder to measure accurately than cluster masses. Still, the high abundance of small voids compared to the most extreme clusters, as well as the high volume filling fraction of voids offers hope that void-galaxy abundances could be used fruitfully.

The dotted curves in the right-hand column show by-eye fits of the form
\begin{align}
\rho_h &= f\rho_m^\alpha \exp\left[\left(\rho_m/\rho_{\rm exp}\right)^{-\varepsilon}\right];\nonumber\\
A_h &= \ln f +\alpha A_m + \exp\left[-\varepsilon \left(A_m-A_{\rm exp}\right)\right].
\label{eqn:adhocfit}
\end{align}
where $\rho\equiv1+\delta$, $A\equiv \ln\rho$, and $f$ is a constant.
Without the exponential factor, this form is $\rho_h=f\rho_m^\alpha$ \citep[e.g.][]{KitauraEtal2014}. In the higher halo-mass bins, high-density pixels do scatter substantially down, which was neglected in making the fits.  We also show vertical one-$\sigma$ error band away from the fits, assuming Poisson statistics in $\delta_h$.

\begin{figure}
    \begin{center}
      \includegraphics[width=\columnwidth]{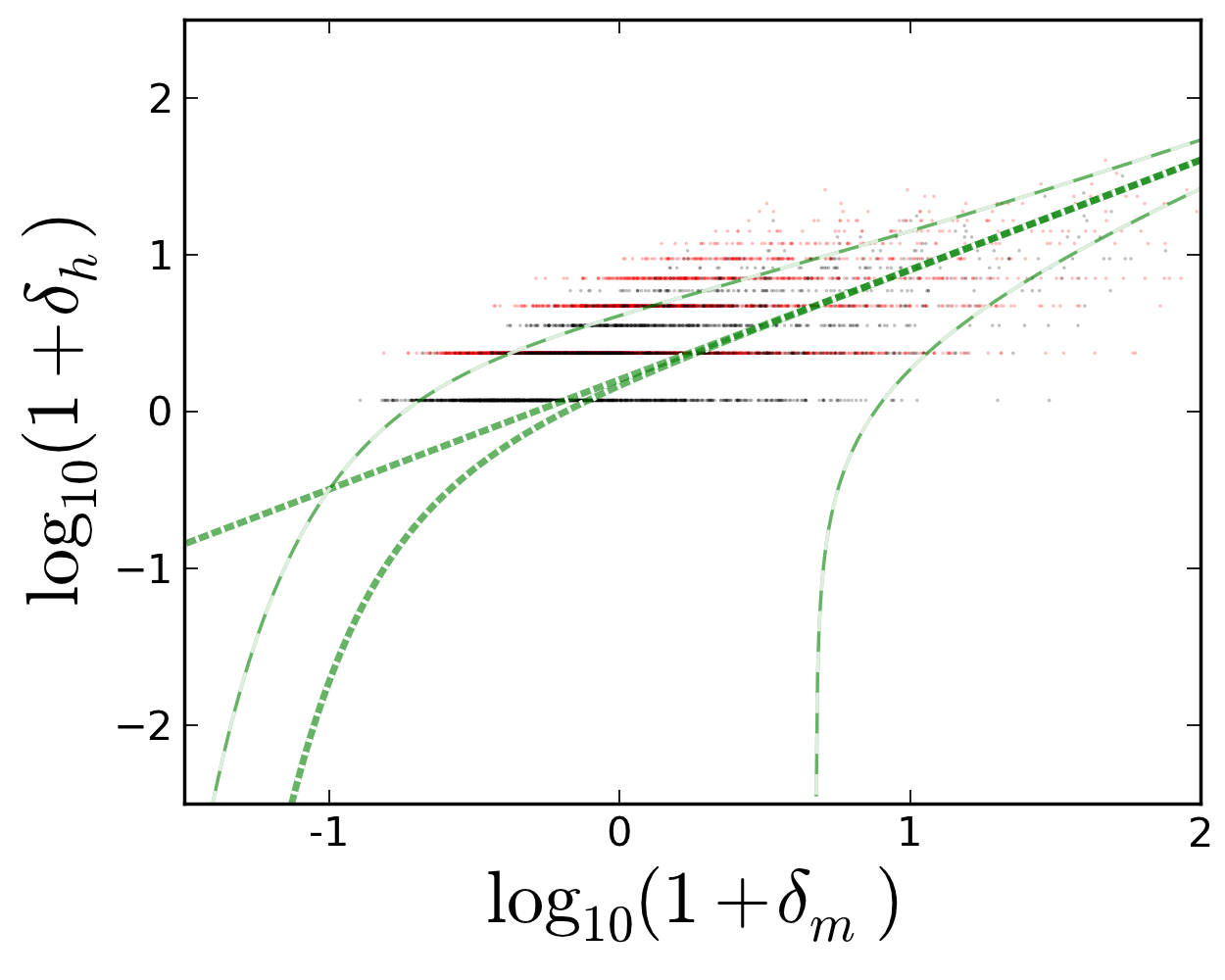}
    \end{center}
        \caption{A scatter plot from a single MIP realization, for $10^{10\mhyphen11} M_\odot/h$ haloes (to be compared with the upper-right panel of Fig.\ \ref{fig:biasscatter}).  The dotted green fit is the same as in Fig.\ \ref{fig:biasscatter}, but the Poisson scatter is much higher, as the large distance between the dotted green curve and the vertical error band around it (dashed curves) indicates. The discreteness here obscures the behavior at low densities.
    \label{fig:biasscatter1}
    }
\end{figure}

To demonstrate the statistical power and low discreteness level of the MIP ensemble average, we show in Fig.\ \ref{fig:biasscatter1} a scatter-plot measured from a single realization. It uses the most-abundant $10^{10\mhyphen11} M_\odot/h$ halo bin, to be compared with the upper-right panel of Fig.\ \ref{fig:biasscatter}. In a single realization, the minimum nonzero halo number density in a cell of volume $V$ is $1/V$, whereas in the MIP ensemble-mean, the minimum nonzero halo number density is $1/(225 V)$.  This high discreteness noise shows up as much-inflated vertical one-$\sigma$ error bands compared to Fig.\ \ref{fig:biasscatter}. With the poor sampling in a single realization, we get barely a hint of the low-density downturn. 

\section{The Poisson approximation}
\label{sec:poissonity}
To model observable, discrete halo samples in the Universe, it is important to know statistics of the point process that produces haloes from the `continuous,' ensemble-mean $\delta_h$.

A common assumption is that the continuous halo-density field is Poisson-sampled in each pixel.  This is an inhomogeneous Poisson process, in which the mean intensity varies with position. At least at the highest densities, various authors have inferred `super-Poisson scatter' in this sampling \citep[e.g.][]{KitauraEtal2014}.  This means that the variance exceeds the mean, unlike in a Poisson process, in which the variance equals the mean. They found that super-Poisson scatter in high-density pixels was necessary to include to model the halo power spectrum accurately at small scales.  The MIP ensemble allows the Poissonity assumption to be tested at low densities.

\begin{figure}
    \begin{center}
      \leavevmode
      \includegraphics[width=\columnwidth]{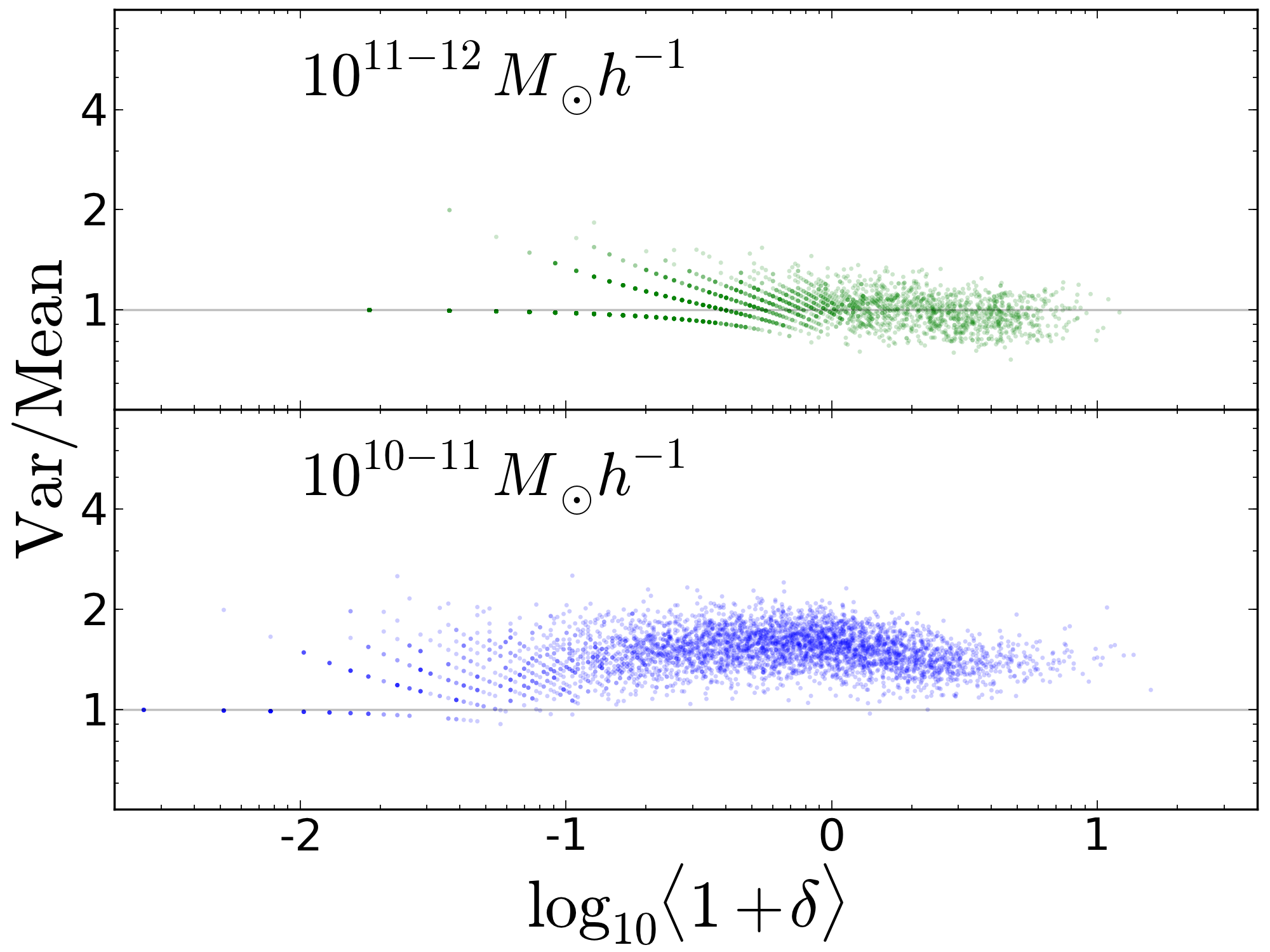}
    \end{center}
    \caption[1]{ \small Scatter plots of the variance-to-mean ratio versus mean halo number densities, across all simulations in the MIP ensemble, for all 2-\hmpc\ pixels containing at least one halo across the ensemble. The faint gray lines correspond to the Poisson ${\rm Mean=Var}$.
    \label{fig:halopoissonity}
    }
\end{figure}

Fig.\ \ref{fig:halopoissonity} is a scatter plot, one dot per 2-\hmpcnosp\ pixel, of $(\sigma^2/\mu)$ against $\mu$, where $\sigma^2$ and $\mu$ are the variance and mean halo densities in each pixel, across MIP simulations. Super-Poissonity (variance exceeding the mean) is prevalent throughout the density range in the lower halo-mass bin.  However, curiously, in the higher-mass bin, Poissonity seems to be a rather good assumption on average.  Recall, though, that this variance does not include the stochasticity from environmental factors beyond the $\delta_h(\delta_m)$ mapping, which we preliminarily investigate in the next section. Note that the ratio here loses meaning when there are only a couple of haloes in the pixel across all simulations. For example, if there is only one halo, this ratio is exactly Poisson because the halo can only belong to one simulation. The discrete allotment of haloes to simulations is the reason for the low-density patterns in each plot. 

\begin{figure}
    \begin{center}
      \leavevmode
      \includegraphics[width=\columnwidth]{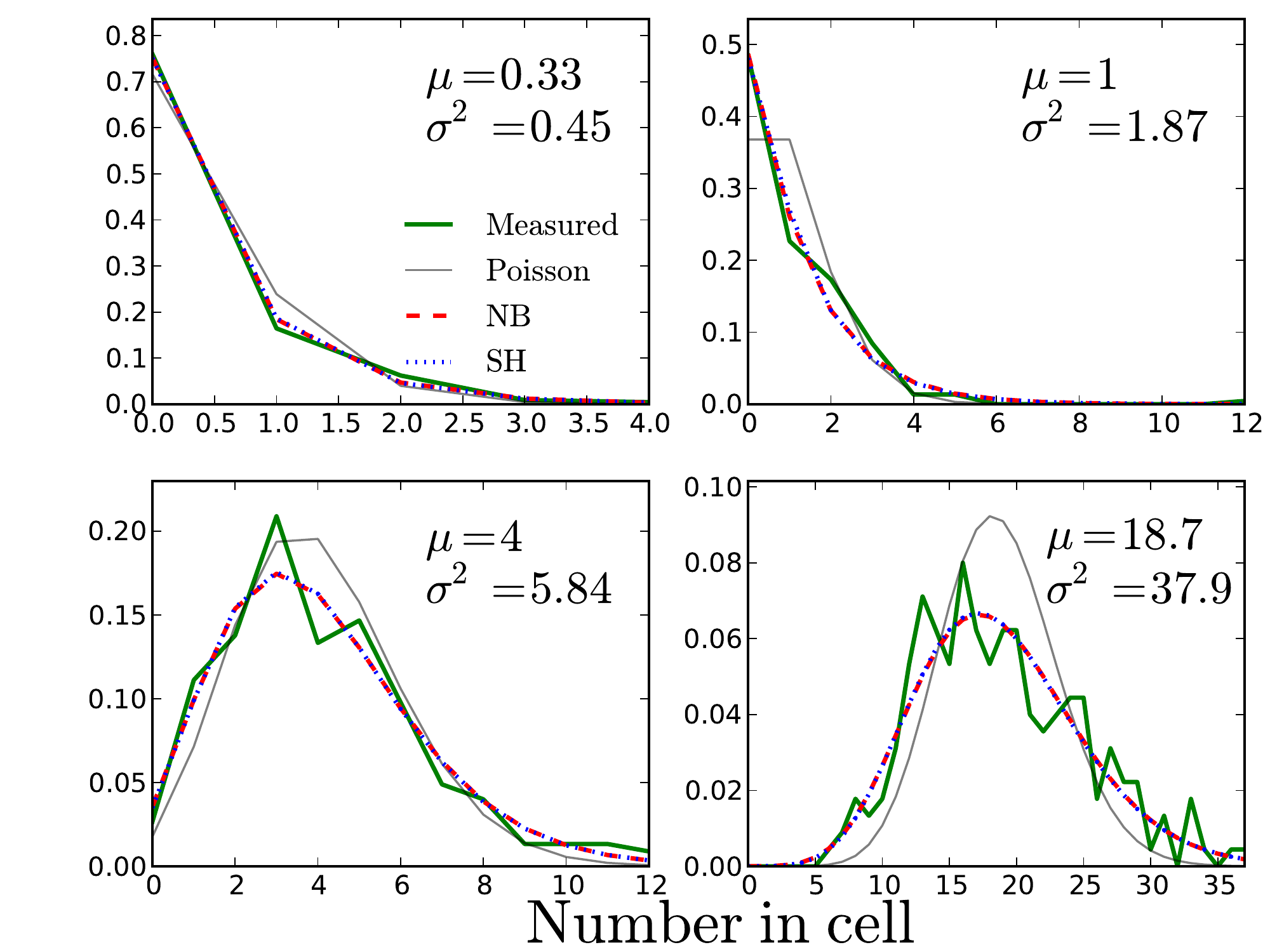}
    \end{center}
    \caption[1]{ \small PDFs across the MIP ensemble of the number of $10^{10\mhyphen11}M_\odot h^{-1}$ haloes in single cells of different mean halo densities. The PDFs (thick solid green) are wider than a Poisson of the same mean (thin solid grey). Also shown are the Saslaw-Hamilton (SH, dotted blue) and negative-binomial (NB, dashed red) distributions that model the super-Poissonity; both look like good fits.  The PDF means and variances are written in each panel.
        \label{fig:poissonitypdfs}
    }
\end{figure}

Fig.\ \ref{fig:poissonitypdfs} goes beyond the mean and variance, and shows the full PDFs across the MIP ensemble in single 2-\hmpcnosp\ cells of different mean halo number densities.  As the plots show, the PDFs in each cell are rather well-fitted with both Saslaw-Hamilton \citep[SH, ][]{SaslawHamilton1984,HamiltonEtal1985} and similarly-shaped negative-binomial (NB) distributions \citep{Sheth1995}. \citet{KitauraEtal2014} successfully use a pixel-by-pixel NB distribution to model the super-Poissonity in high-density pixels; based on our results, this seems to be a good strategy. The SH and NB distributions are as follows:
\begin{align}
 f_{\rm SH}(N) & = \frac{\lambda}{N!}e^{-\lambda(1-b)-Nb}(1-b)[\lambda(1-b)+Nb)]^{N-1}; \\
 f_{\rm NB}(N) & = \frac{\lambda}{N!}\frac{\Gamma(\beta+N)}{\Gamma(\beta)(\beta+\lambda)^N (1+\lambda/\beta)^\beta}.
 \label{eqn:fshnz}
\end{align}
Here $\lambda$ is the mean, and the variance $\sigma^2$ determines the parameters $b=1-\sqrt{\lambda/\sigma^2}$, and $\beta=\lambda^2/(\sigma^2-\lambda)$. An NB distribution can arise in a process in which the Poisson mean parameter $\lambda$ is itself a gamma-distributed (similar to lognormal-distributed) random variable. That is, our results are consistent with a gamma (or, perhaps, lognormal) distribution of $1+\delta_m$ in each cell, which is then Poisson-sampled.

\section{Stochasticity in the continuous halo-density field}
\label{sec:cosmicweb}
While the scatter plots in Fig.\ \ref{fig:biasscatter} are rather tight, there is substantial scatter at high $\delta_m$. Especially for high-mass halos, this scatter seems to be beyond the scatter from the point process in each individual pixel, i.e.\ the error bands in Fig.\ \ref{fig:biasscatter}. The scatter in $\delta_h$ for different pixels of the same $\delta_m$ relates to elements of the pixel `environment' beyond the density measured in 2 or 4\hmpc\ cells.

\begin{figure}[H]
    \begin{center}
      \leavevmode
      \includegraphics[width=\columnwidth]{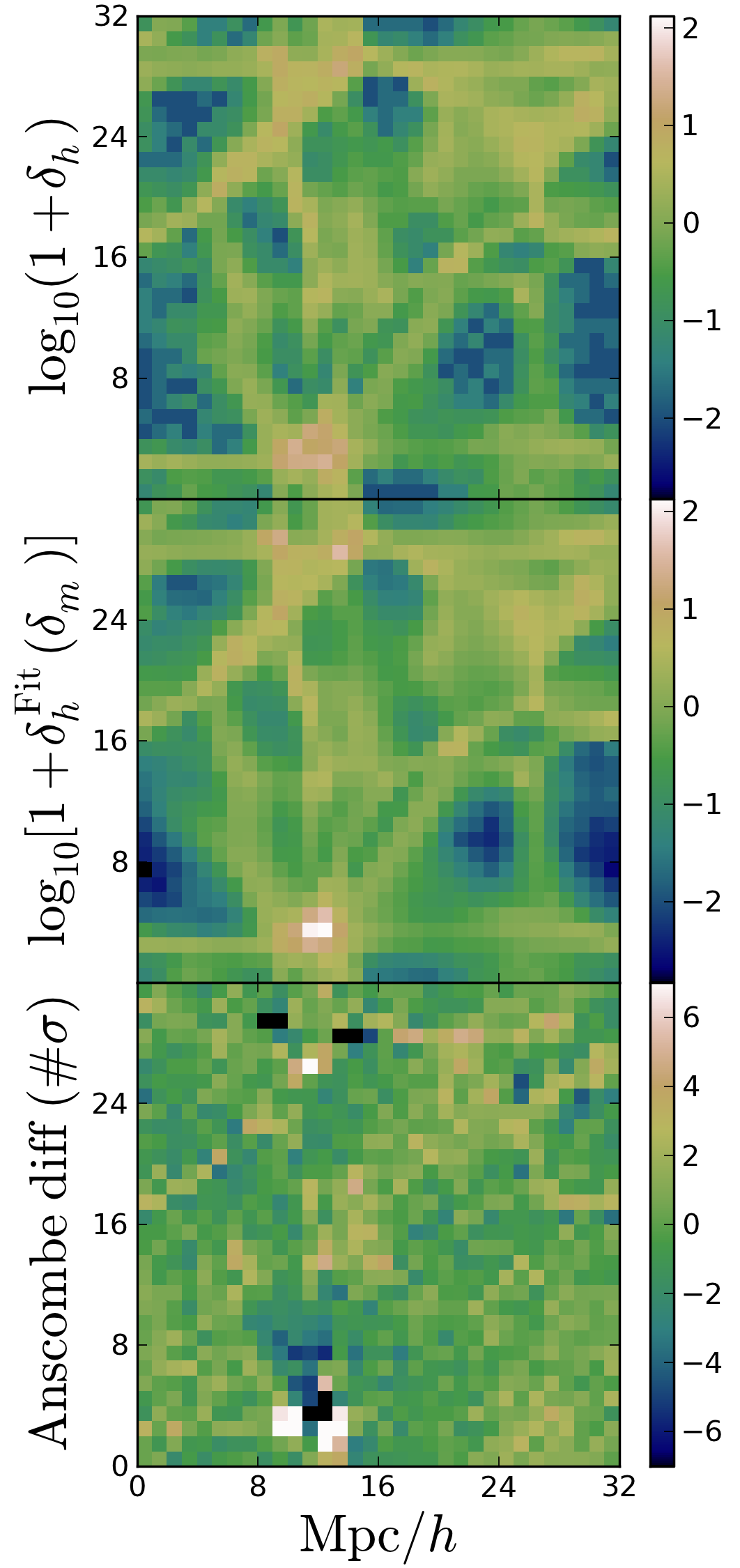}
    \end{center}
    \caption[1]{ \small A comparison of (top) the ensemble-mean $10^{10\mhyphen11} M_\odot/h$ halo log-density field, and (middle) as predicted from the matter log-density field according to the fit in the middle-right panel of Fig.\ \ref{fig:biasscatter}. In the top panel, the log-density for zero-density pixels is set as though there were half a halo in the cell. The bottom panel shows the Anscombe-transformed difference between the two (see text). This is designed to show an estimate of the number of standard deviations that the top panel is away from the middle panel, assuming independent Poisson statistics in each pixel.
        \label{fig:cosmicwebcomp}
    }
\end{figure}

As a preliminary test of whether this stochasticity is caused by visually evident factors such as the cosmic-web environment, we show in Fig.\ \ref{fig:cosmicwebcomp} a two-dimensional slice of MIP ensemble-mean densities, with 1\hmpc\ pixels. This slice contains the highest-$\delta_m$ pixel in the MIP, the cluster at bottom. For haloes of mass $10^{11\mhyphen12} M_\odot/h$, Fig.\ \ref{fig:cosmicwebcomp} shows log-densities of three fields: the actual ensemble-mean halo-density field; the halo-density field as predicted from the matter-density field using the empirical fit (Eq.\ \ref{eqn:adhocfit}) in the upper-right panel of Fig.\ \ref{fig:biasscatter}; and in the bottom panel, we plot the `Anscombe difference' between the two, an estimate of the number of standard deviations away from the expected mean halo density in each cell, assuming Poisson statistics.

In the bottom panel, we do not show the raw difference in log-densities, since this would mostly show the differences in voids, where both the increased discreteness and the density downturn make the fractional noise large. Instead, we use an \citet{Anscombe1948} transform separately in each cell, which is designed to transform Poisson-distributed data into Gaussian-distributed data. The Anscombe transform is ${\rm Ansc}(x)=2\sqrt{x+3/8}$. A Poisson-distributed variable of mean $\lambda$ Anscombe-transforms into a Gaussian of mean $\mu_{\rm Ansc}(\lambda)=2\sqrt{\lambda+3/8}-1/(4\sqrt{\lambda})$, and variance 1. Explicitly, what we show is ${\rm Ansc}(N_h)-\mu_{\rm Ansc}[\lambda=\bar{N}_h(1+\delta_h(\delta_m))]$, where $N_h$ is the number of haloes in the cell in the MIP ensemble stack, and $\bar{N}_h$ is the mean number of haloes.

Visually, the level of Anscombe-normalized fluctuation seems to increase with density, growing large in filaments, and larger still in clusters. There seems to be a tendency for filaments to have a higher $\delta_h$ than $\delta_m$ would predict, at least according to our fit. At least in high-density regions, $\delta_h$ deviates much more than expected in a Gaussian (Anscombe-transformed Poisson) distribution, another sign that the point process is super-Poisson.

The cluster at the bottom shows huge deviations in halo density away from that predicted by the matter, consistent with the substantial overdispersion of the halo density compared to Poisson in high-density areas. Morphologically, the cluster center is underpopulated with haloes compared to what $\delta_m$ would predict, and its outskirts are overpopulated. We suspect that this is because this plot is made from rather low-mass haloes, while most of the mass in the cluster would likely go into the high-mass haloes (or perhaps a single large halo) in the cluster center. Including all haloes, and weighting them by their mass as \citet{ParkEtal2010} do, might reduce this effect.  This effect may also come from slight shifts in the cluster's position with different sets of small-scale modes.

Fig.\ \ref{fig:cosmicwebcomp} visually indicates that filaments and large clusters do have some influence on $\delta_h$; in filaments, $\delta_h$ tends to be larger, and in clusters, the cores tend to be more centrally-concentrated in matter than in haloes. This suggests that in high-density regions (typically either filament or cluster environments), at fixed $\delta_m$, $\delta_h$ is higher in filaments than in clusters.  These differences may come from influences of the tidal field, or perhaps halo exclusion in clusters. However, comparing the top and middle panels, it is clear that $\delta_m$ itself is by far the dominant factor determining $\delta_h$.  This analysis is currently inconclusive, however; we plan to revisit this issue in future work, incorporating a quantitative cosmic-web classification.

\section{Gaussianization: a test of stochasticity}
\label{sec:gaussianization}
Another way to test for systematic fluctuations in the ensemble-mean relation between $\delta_h$ and $\delta_m$ is to compare their power spectra and Fourier-space cross correlations, which we do in this section. Suppose that $\delta_h$ is a strictly increasing local function of $\delta_m$. Then, if both are mapped to give the same PDF, they will be the same fields. A natural choice of PDF to map each field onto is a Gaussian \citep{Weinberg1992}, since, for instance, the power spectrum of a field after this (rank-order) Gaussianization has low covariance \citep{NeyrinckEtal2009}. This benefit of a Gaussianized field, that the result is insensitive to any monotonic transformation made on the field before it is Gaussianized, has long been exploited for topological statistics such as the genus \citep[e.g.][]{WeinbergEtal1987}.

Denote as ${\rm Gauss}(\delta)$ the field $\delta$ after Gaussianization.  Explicitly, if $\delta$ is a field defined on a finite number of pixels $N$, ${\rm Gauss}(\delta) =\sqrt{2}\sigma {\rm erf}^{-1}(2f_{<\delta}-1+1/N)$, where $f_{<\delta}$ is the fraction of pixels less-dense than $\delta$, and $\sigma$ is the standard deviation of the Gaussian that $\delta$'s PDF is mapped onto. If the function $\delta_h(\delta_m)$ is strictly increasing, then $f_{<\delta_h(\delta_m)} = f_{<\delta_m}$, so ${\rm Gauss}(\delta_h)={\rm Gauss}(\delta_m)$.

\begin{figure}
    \begin{center}
      \leavevmode
      \includegraphics[width=\columnwidth]{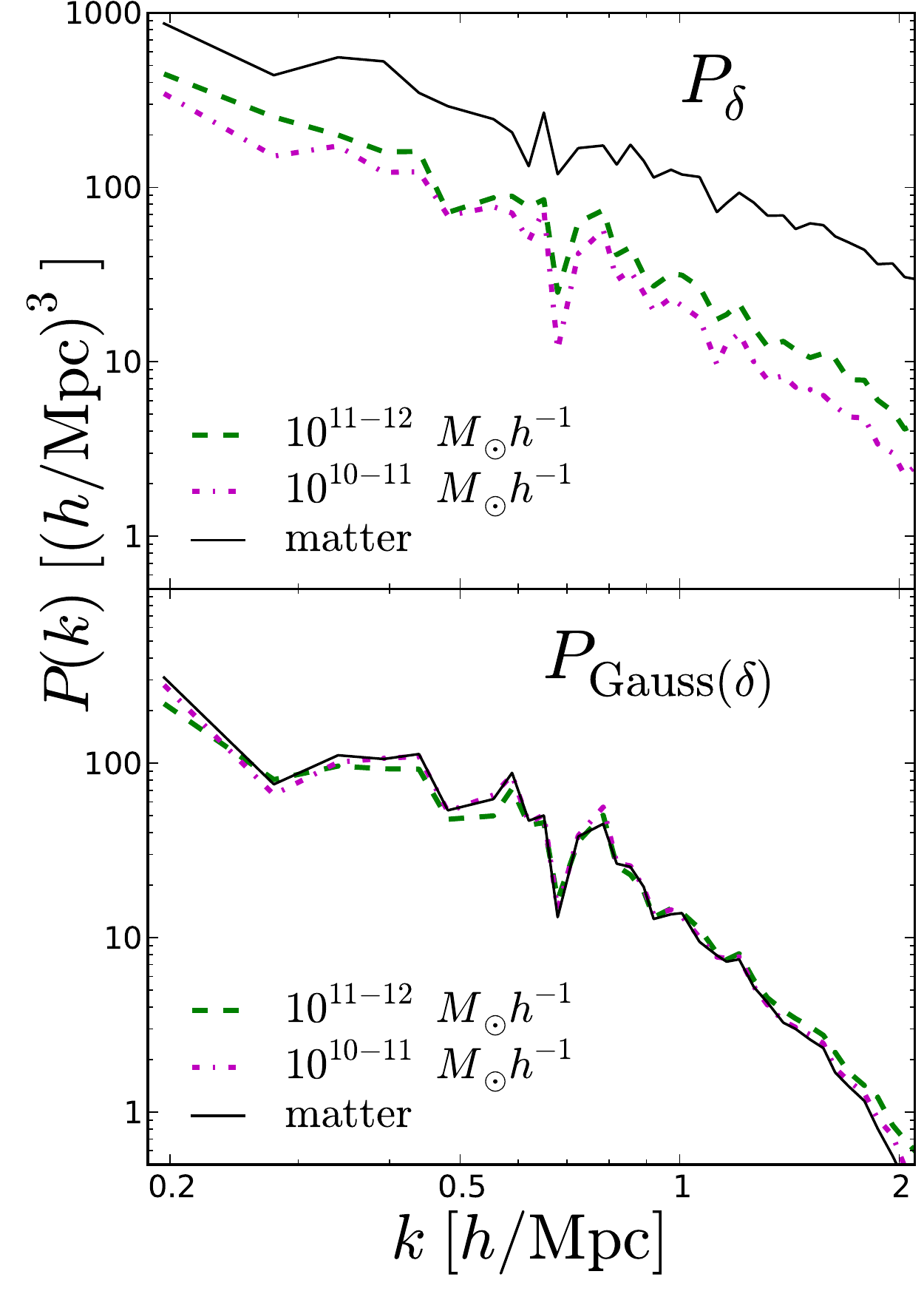}
    \end{center}
    \caption[1]{ \small Power spectra of matter and three mass ranges of haloes in the MIP ensemble-mean fields. The Gaussianized-density power spectra $P_{{\rm Gauss}(\delta)}$ show substantially less difference among the various density fields than the raw density power spectra $P_\delta$, supporting the hypothesis that a local, strictly-increasing density mapping captures the mean relationship between matter and haloes.
    \label{fig:gausspower}
    }
  \end{figure}

Fig.\ \ref{fig:gausspower} shows power spectra of both $\delta$ and ${\rm Gauss}(\delta)$, with Gaussianization done on a 2-\hmpcnosp\ grid, for both the ensemble-mean matter and halo-density fields.  If the power spectrum were shown from a single simulation, it would have a noticeable shot noise, but this discreteness is negligible in the ensemble mean. While each $P_\delta$ varies substantially from each other, the Gaussianized power spectra are much more similar, the halo curves only substantially departing from the matter curve for haloes in the largest-mass bin.  The amplitude of each curve is the same because the standard deviation $\sigma$ used for Gaussianization is set to 1 in each case.

\begin{figure}
    \begin{center}
      \leavevmode
      \includegraphics[width=\columnwidth]{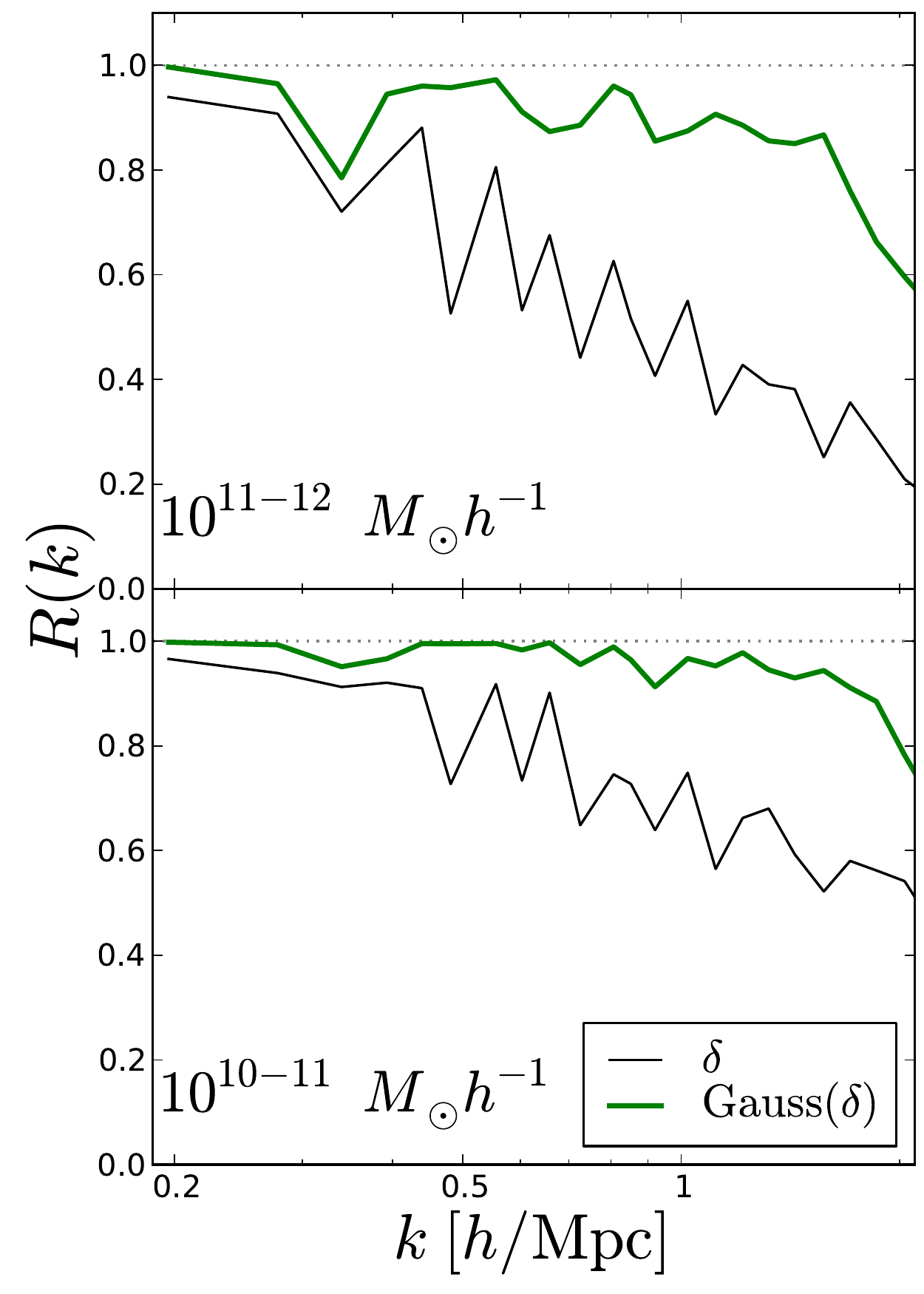}
    \end{center}
    \caption[1]{ \small Fourier-space cross-correlations of spectra of matter and three mass ranges of haloes in the MIP ensemble-mean fields. As with the power spectra, the halo density fields show substantially less difference from the matter density fields if both are Gaussianized, supporting the hypothesis that a local, strictly-increasing density transform largely captures the mean relationship between matter and haloes.
    \label{fig:gausscross}
    }
\end{figure}

Fig.\ \ref{fig:gausscross} shows a common measure of stochasticity, the cross-correlation coefficient $R(k)=P_{h\times m}(k)/\sqrt{P_h(k) P_m(k)}$, where $P_{h\times m}$ is the halo-matter cross spectrum.  To ensure that $|R(k)| \le 1$ \citep{NeyrinckEtal2005}, as one might hope from the Schwarz inequality, we do not subtract the (small, in our case) shot noise from $P_h$ when forming this ratio.  $R(k)$ is sensitive to each individual Fourier amplitude and phase, while $P(k)$ is sensitive only to Fourier amplitudes, as averaged within bins.  Indeed, $R(k)$ is closer to unity (that is, the halo and matter fields are more similar) for each halo field if both the halo and matter fields are Gaussianized.

This is an example of the superiority of the statistics of Gaussianized fields.  $R(k)$ measured from the raw $\delta$ fields is a measure of stochasticity, invariant to a linear bias between haloes and matter. But the Fourier cross-correlation of the Gaussianized fields has even greater power as a measure of stochasticity, because it is invariant not only to a linear bias, but to any strictly-increasing biasing function.

While these results show that to an excellent approximation, the ensemble-mean halo-density fields are local, strictly-increasing transformations of the matter-density field, it is important to remember that an observed galaxy-density field would typically have substantial discreteness and halo-exclusion effects, that are negligible here.  The influence of these effects on Gaussianized fields can be straightforwardly measured using the MIP ensemble, but we leave their full investigation to future work. 

\section{Conclusions}
In this paper, we have measured the bias relation giving the mean halo density in terms of the dark-matter density down to unprecedentedly low densities.  The new tool that made this possible was the MIP ensemble, in which a large-scale cosmic web is sampled in hundreds of ways by changing initial small-scale modes underneath fixed large-scale modes. The ensemble-mean we measure from both fields has negligible stochasticity and exclusion; it is instructive to separate out these from effects of the large-scale cosmic-web environment.

The form we find for the bias relation is that of a power law and an exponential at low densities, which are fit well with two models; one is a standard excursion-set model, and another is a new `local growth factor' model, in which small-scale modes grow as though they were in a homogeneous universe with $\Omega_m$ proportional to the local large-scale density. These models imply that the abundance of modest-size haloes in voids is quite sensitive to cosmological parameters.

The strictly-increasing bias function giving $\delta_h$ from $\delta_m$ is promising for cosmological tests involving voids, which often assume that voids found in a galaxy sample correspond to dark-matter voids.  In addition, the tight relationship we found suggests that the bias function can be used to directly convert a void profile found in haloes to a matter void profile.  This remains to be explicitly shown, since exclusion and discreteness could corrupt the relationship, but this test would be straightforward using the MIP ensemble. 

A tight $\delta_h$-$\delta_m$ relationship also implies a similarity in their power spectra if a local transform is done on one or both to give the same PDF. A statistically convenient PDF is a Gaussian; we check that indeed, after rank-order Gaussianizing the ensemble-mean fields, they have high cross-correlation and impressively similar power spectra for  two halo-mass ranges.

There are still many aspects of the dark-matter-to-halo mapping to explore, all of which the MIP ensemble can help to study. The current investigation is entirely in real space, but it is essential to investigate redshift space, as well; for example, it is possible that in fingers of God, haloes sample the matter in a different way than in real space. The assumption that the halo density field is a \mbox{(super-)Poisson} sampling of a continuous field also should be investigated in more detail; for example, in the sampling process, substantial covariance among nearby pixels could occur due to halo exclusion. The effects of discreteness on statistics like the power spectrum, and especially the Gaussianized- or log-density power spectrum, would also benefit from study with the MIP ensemble.

\section*{Acknowledgements}
We thank Francisco-Shu Kitaura, Ravi Sheth, Ixandra Achitouv, Aseem Paranjape, and an unusually helpful anonymous referee for useful comments. MCN and MAAC are grateful for support from a New Frontiers in Astronomy and Cosmology grant from the Sir John Templeton Foundation. DJ is supported by DoE SC-0008108, NASA NNX12AE86G, and NSF 0244990.

\bibliography{refs}

\section*{Appendix}
\begin{figure}
    \begin{center}
      \leavevmode
      \includegraphics[width=\columnwidth]{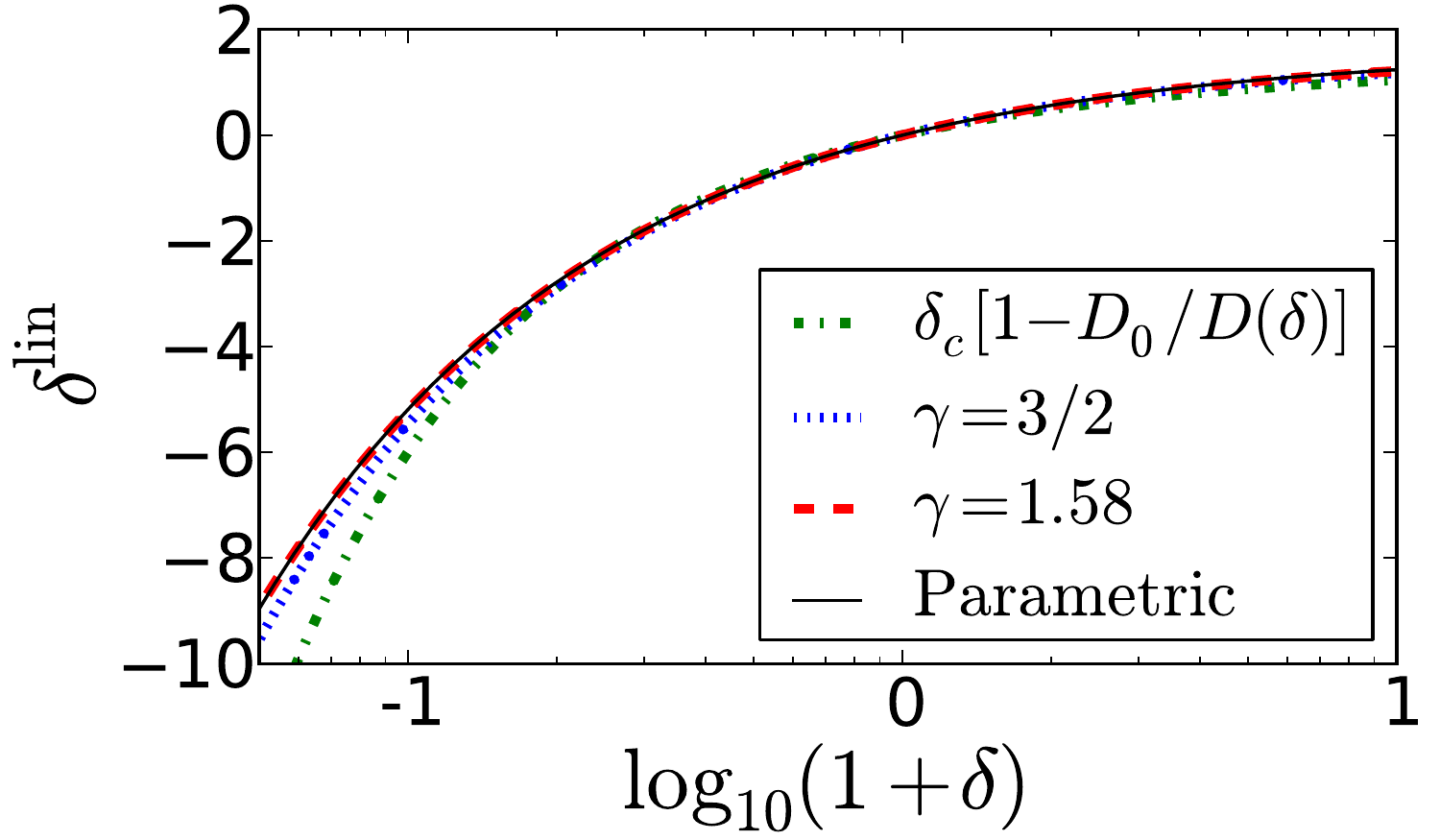}
    \end{center}
    \caption[1]{ \small Performance of a few approximations to the parametric spherical-collapse relation between initial ($\delta^{\rm lin}$) and final non-linear density $\delta$.
    \label{fig:testsphercol}
    }
\end{figure}

The agreement between the AES and LGF models in \S \ref{sec:mattertohaloes} implies that subtracting a linear-theory large-scale density ($\deltallin$) from $\deltac$ is roughly equivalent to multiplicatively scaling $\deltac$ by $D_0/D(\deltal)$. Again, as \citet{MartinoSheth2009} found, a greater equivalence would result if an effective change to $\Omega_\Lambda$ is implemented; here we test the equivalence without this.

Setting the arguments in the exponentials of Eqs.\ (\ref{eqn:aes}) and (\ref{eqn:lgf}) equal, 
\begin{equation}
\deltallin=\deltac[1-D_0/D(\deltal)].
\label{eqn:newdelta}
\end{equation}
Fig.\ \ref{fig:testsphercol} shows how this formula compares to the parametric solution used for the results in the paper \citep{ShethvandeWeygaert2003}. It also compares two `local-Lagrangian' models \citep{ProtogerosScherrer1997}, that contain a parameter $\gamma$ that equals $3/2$ in a low-$\Omega_m$, $\Omega_\Lambda=0$ limit of the spherical-collapse model \citep{Bernardeau1992}, that works quite well for voids \citep[e.g.][]{Neyrinck2013}.  The local-Lagrangian parameterization is
\begin{equation}
\deltallin=\gamma[1-(1+\deltal)^{-1/\gamma}].
\label{eqn:locallag}
\end{equation}

The new model involving the growth factor does not perform badly for $1+\delta>0.1$, but the local-Lagrangian models definitely work better. Curiously, $\gamma=1.58$, between 3/2 and $\deltac=1.686$, works best, giving a curve nearly indistinguishable from the parametric solution.

\end{document}